%% file: skeleton.tex
\documentclass{PoS}
\usepackage[utf8]{inputenc}
\usepackage{amsmath,xspace}
\newcommand\url[1]{\href{#1}{\tt #1}}
\usepackage[normalem]{ulem} %%% ==> To implement corrections, to be deleted once it is done

\title{State of the art POWHEG generators for top mass measurements at the LHC}

\ShortTitle{State of the art POWHEG generators for top mass measurements at the LHC}

\author{Silvia Ferrario Ravasio\\
        Universit\`a di Milano-Bicocca and INFN, Sezione di Milano-Bicocca,
        Piazza della Scienza 3, 20126 Milano, Italy\\
        E-mail: \email{silvia.ferrario@mib.infn.it}}

\author{\speaker{Tom\'a\v{s} Je\v{z}o}\thanks{Work supported by the Swiss National Science Foundation~(SNF) under
contracts BSCGI0-157722 and CRSII2-160814.}\\
        Physics Institute, Universit\"at Z\"urich, Z\"urich, Switzerland\\
        E-mail: \email{tomas.jezo@physik.uzh.ch}}

\input ./texmacros.tex

\input ./global_definitions.tex

\abstract{
We study the theoretical uncertainties in the determination of the top-quark mass using next-to-leading-order~(NLO) generators, that describe the top-quark decay at different levels of accuracy, interfaced to parton showers~(PS).
Specifically we consider one generator that implements NLO corrections only in the production dynamics, one that also takes them into account in the top-quark decay in the narrow width approximation (NWA) and one that implements them exactly, including finite-width and interference effects.
We aim at assessing the errors in top-mass determinations of purely theoretical origin.
We do so by measuring relative peak position shifts of $Wb$-jet mass distributions.
Besides the theoretical errors due to the use of less accurate NLO+PS generators, we also explore uncertainties related to shower and modelling of non-perturbative effects by comparing the results obtained by interfacing our generators to both \Pythia{} and \Herwig{} shower Monte Carlos~(SMCs).
}

\FullConference{XXVI International Workshop on Deep-Inelastic Scattering and Related Subjects (DIS2018)\\
		16-20 April 2018\\
		Kobe, Japan}

\begin{document}

\section{Introduction}
%%%%%%%%%%%%%%%%%%%%%%%%%%%%%%%%%%%%%%%%%%%%%%%%%%%%%%%%%%%%%%%%%%%%%%%%%%%%%%%%%%%%%%%%%%%%%%%%%%%%%%%%%%%%%%%%%%%%%%%%%%%%%%%%%%%%%%%%%%%%%%%%%%%%
The abundant production of top pairs at the Large Hadron Collider~(LHC) provides an opportunity for detailed studies of top-quark properties, for tests of the Standard Model~(SM) in the top sector, and for measurements of fundamental parameters such as the top-quark mass.
With the Higgs boson mass now known with high precision, the $W$-boson and top-quark masses have become strongly correlated, and an accurate determination of both would lead to a SM test of
unprecedented precision~\cite{Patrignani:2016xqp-EWreview, Baak:2014ora}.
%%%%%%%%%%%%%%%%%%%%%%%%%%%%%%%%%%%%%%%%%%%%%%%%%%%%%%%%%%%%%%%%%%%%%%%%%%%%%%%%%%%%%%%%%%%%%%%%%%%%%%%%%%%%%%%%%%%%%%%%%%%%%%%%%%%%%%%%%%%%%%%%%%%%

%%%%%%%%%%%%%%%%%%%%%%%%%%%%%%%%%%%%%%%%%%%%%%%%%%%%%%%%%%%%%%%%%%%%%%%%%%%%%%%%%%%%%%%%%%%%%%%%%%%%%%%%%%%%%%%%%%%%%%%%%%%%%%%%%%%%%%%%%%%%%%%%%%%%
Many top-mass measurements are performed by fitting $\mt$-dependent kinematic distributions to Monte Carlo predictions. 
The most precise ones, generally referred to as the \emph{direct measurements}, rely upon a full or a partial reconstruction of the system of the top-decay products.
The CMS and ATLAS measurements of Refs.~\cite{Khachatryan:2015hba} and~\cite{Aaboud:2016igd}, yielding the value $\mt=172.44 \pm 0.13~{\rm (stat)} \pm 0.47$~(syst)~GeV and $172.84 \pm 0.34~{\rm (stat)} \pm 0.61$~(syst)~GeV respectively, fall into this broad category.  
By extracting the $\mt$ value from a fit to the measured distributions, we are unavoidably affected by theoretical errors that must be carefully assessed.
The errors will depend upon the accuracy of the modelling of these distributions.
%%%%%%%%%%%%%%%%%%%%%%%%%%%%%%%%%%%%%%%%%%%%%%%%%%%%%%%%%%%%%%%%%%%%%%%%%%%%%%%%%%%%%%%%%%%%%%%%%%%%%%%%%%%%%%%%%%%%%%%%%%%%%%%%%%%%%%%%%%%%%%%%%%%%

%%%%%%%%%%%%%%%%%%%%%%%%%%%%%%%%%%%%%%%%%%%%%%%%%%%%%%%%%%%%%%%%%%%%%%%%%%%%%%%%%%%%%%%%%%%%%%%%%%%%%%%%%%%%%%%%%%%%%%%%%%%%%%%%%%%%%%%%%%%%%%%%%%%%%
The availability of two new \POWHEGBOX{}~\cite{Nason:2004rx, Frixione:2007vw, Alioli:2010xd} generators for $t\bar{t}$ production, \ttbnlodec{}~\cite{Campbell:2014kua} and \bbfourl{}~\cite{Jezo:2016ujg}, provides us with an opportunity to explore theoretical uncertainties in the top-mass determination that are customarily not considered in experimental analyses.
In particular we are in a position to assess whether NLO corrections in top decay, implemented in both \ttbnlodec{} and \bbfourl{}, and finite width effects, non-resonant contributions and interference of radiation from production and decay, implemented in \bbfourl{}, can lead to sizeable corrections to the extracted value of the top mass. 
Since the \hvq{} generator~\cite{Frixione:2007nw} that implements NLO corrections only in production is widely used by the experiments in top-mass analyses, we are particularly interested in assessing to what extent it is compatible with them. 
The impact of NLO corrections in decays and finite-width effects in top-mass determinations at fixed order were considered in Refs.~\cite{Heinrich:2017bqp, Bevilacqua:2017ipv}.
%%%%%%%%%%%%%%%%%%%%%%%%%%%%%%%%%%%%%%%%%%%%%%%%%%%%%%%%%%%%%%%%%%%%%%%%%%%%%%%%%%%%%%%%%%%%%%%%%%%%%%%%%%%%%%%%%%%%%%%%%%%%%%%%%%%%%%%%%%%%%%%%%%%%%

%%%%%%%%%%%%%%%%%%%%%%%%%%%%%%%%%%%%%%%%%%%%%%%%%%%%%%%%%%%%%%%%%%%%%%%%%%%%%%%%%%%%%%%%%%%%%%%%%%%%%%%%%%%%%%%%%%%%%%%%%%%%%%%%%%%%%%%%%%%%%%%%%%%%
We focus on NLO+PS and matching related effects potentially important in top-mass determinations from direct measurements.
As an observable we chose the peak position of the reconstructed mass of the top-quark, that we define as a system comprising a hard lepton, a hard neutrino and a hard $b$ jet.
We consider it both at the particle truth level, and by including experimental inaccuracies, that we assume can be implemented through a simple smearing with a resolution function, a Gaussian with a width of 15~GeV, which is the typical resolution achieved on the top mass by the LHC collaborations.
Firstly, we are interested in understanding to what extent the peak position depends on the chosen NLO+PS generator.
Such dependence would be evidence that the new features introduced in more modern generators are mandatory for accurate mass extractions.
Furthermore, we also attempt to give a first assessment of ambiguities associated with shower and non-perturbative effects by interfacing our NLO+PS generators to several shower SMC programs.
%%%%%%%%%%%%%%%%%%%%%%%%%%%%%%%%%%%%%%%%%%%%%%%%%%%%%%%%%%%%%%%%%%%%%%%%%%%%%%%%%%%%%%%%%%%%%%%%%%%%%%%%%%%%%%%%%%%%%%%%%%%%%%%%%%%%%%%%%%%%%%%%%%%%

%%%%%%%%%%%%%%%%%%%%%%%%%%%%%%%%%%%%%%%%%%%%%%%%%%%%%%%%%%%%%%%%%%%%%%%%%%%%%%%%%%%%%%%%%%%%%%%%%%%%%%%%%%%%%%%%%%%%%%%%%%%%%%%%%%%%%%%%%%%%%%%%%%%%
In these proceedings we briefly review some of the findings of a much more extensive and complete study of theoretical uncertainties in top-mass determinations performed in Ref.~\cite{Ravasio:2018lzi}, that this manuscript derives from.
We also further explore the ambiguities associated with shower and non-perturbative effects by extending the set of SMCs we interface to, including the original \PythiaEightPtwo{}~\cite{Sjostrand:2014zea} and \HerwigSevenPone{}~\cite{Bahr:2008pv, Bellm:2015jjp}, and adding their older versions \PythiaSixPfour{}~\cite{Sjostrand:2006za} and \HerwigSixPfive{}~\cite{Corcella:2000bw,Corcella:2002jc}.
These results are new and have not been included in other publications.
%%%%%%%%%%%%%%%%%%%%%%%%%%%%%%%%%%%%%%%%%%%%%%%%%%%%%%%%%%%%%%%%%%%%%%%%%%%%%%%%%%%%%%%%%%%%%%%%%%%%%%%%%%%%%%%%%%%%%%%%%%%%%%%%%%%%%%%%%%%%%%%%%%%%

\section{NLO+PS generators and their interfaces to SMC}
\label{sec:generators}

%%%%%%%%%%%%%%%%%%%%%%%%%%%%%%%%%%%%%%%%%%%%%%%%%%%%%%%%%%%%%%%%%%%%%%%%%%%%%%%%%%%%%%%%%%%%%%%%%%%%%%%%%%%%%%%%%%%%%%%%%%%%%%%%%%%%%%%%%%%%%%%%%%%%
The \hvq{} program~\cite{Frixione:2007nw} is the first top-pair production generator implemented in \POWHEG{}. 
It uses on-shell matrix elements for NLO production of $t\bar{t}$ pairs. 
Off-shell effects and top decays, including spin correlations, are introduced in an approximate way~\cite{Frixione:2007zp}.
Radiation in decays is fully handled by the PS generators, eventually also implementing matrix-element corrections~(MEC).
%%%%%%%%%%%%%%%%%%%%%%%%%%%%%%%%%%%%%%%%%%%%%%%%%%%%%%%%%%%%%%%%%%%%%%%%%%%%%%%%%%%%%%%%%%%%%%%%%%%%%%%%%%%%%%%%%%%%%%%%%%%%%%%%%%%%%%%%%%%%%%%%%%%%

%%%%%%%%%%%%%%%%%%%%%%%%%%%%%%%%%%%%%%%%%%%%%%%%%%%%%%%%%%%%%%%%%%%%%%%%%%%%%%%%%%%%%%%%%%%%%%%%%%%%%%%%%%%%%%%%%%%%%%%%%%%%%%%%%%%%%%%%%%%%%%%%%%%%
In order to handle radiative corrections in top decays the \POWHEG{} method had to be generalized, yielding the improved \ttbnlodec{}~\cite{Campbell:2014kua} generator. 
It describes the $t\bar{t}$ process in the NWA including offshell and non-resonant effects by reweighting with the exact LO matrix elements.
A general procedure for dealing with decaying resonances that can radiate by strong interactions has been introduced and implemented in a fully automatic way in the \RES~\cite{Jezo:2015aia} code.
Besides radiation in resonance decays, this framework allows for the treatment of off-shell effects, non-resonant subprocesses and their quantum interference, and interference of radiation from production and resonance decay.\footnote{A related approach within the \MCatNLO{} framework has been presented in Ref.~\cite{Frederix:2016rdc}.}
Ref.~\cite{Jezo:2016ujg} introduces the \bbfourl{} generator that implements the process $pp\to b\bar{b}\,\fourl$, including all QCD NLO corrections in the 4-flavour scheme, i.e.~accounting for finite $b$-mass effects.
So double-top, single-top and non-resonant diagrams are all included with full spin-correlation effects, radiation in production and decays, and their interference.
%%%%%%%%%%%%%%%%%%%%%%%%%%%%%%%%%%%%%%%%%%%%%%%%%%%%%%%%%%%%%%%%%%%%%%%%%%%%%%%%%%%%%%%%%%%%%%%%%%%%%%%%%%%%%%%%%%%%%%%%%%%%%%%%%%%%%%%%%%%%%%%%%%%%

%%%%%%%%%%%%%%%%%%%%%%%%%%%%%%%%%%%%%%%%%%%%%%%%%%%%%%%%%%%%%%%%%%%%%%%%%%%%%%%%%%%%%%%%%%%%%%%%%%%%%%%%%%%%%%%%%%%%%%%%%%%%%%%%%%%%%%%%%%%%%%%%%%%%
In \RES{} and in \ttbnlodec{}, radiation is generated according to the \POWHEG{} Sudakov form factor both for the production and for all resonance decays that involve coloured partons.  
This feature also offers the opportunity to modify the standard \POWHEG{} single-radiation approach.  
Rather than picking the hardest radiation from one of all possible origins (i.e.~production and resonance decays), the \POWHEGBOX{} can generate simultaneously the hardest radiation in production and in each resonance decay.
Such multiple-radiation events have to be completed by a SMC program, that has to generate radiation from each origin without exceeding the hardness of the corresponding \POWHEG{} one.
This requires an interface that goes beyond the simple Les Houches~(LH) standard~\cite{Boos:2001cv}, since the radiation from decaying resonances is generated unrestricted by default.
%%%%%%%%%%%%%%%%%%%%%%%%%%%%%%%%%%%%%%%%%%%%%%%%%%%%%%%%%%%%%%%%%%%%%%%%%%%%%%%%%%%%%%%%%%%%%%%%%%%%%%%%%%%%%%%%%%%%%%%%%%%%%%%%%%%%%%%%%%%%%%%%%%%%

%%%%%%%%%%%%%%%%%%%%%%%%%%%%%%%%%%%%%%%%%%%%%%%%%%%%%%%%%%%%%%%%%%%%%%%%%%%%%%%%%%%%%%%%%%%%%%%%%%%%%%%%%%%%%%%%%%%%%%%%%%%%%%%%%%%%%%%%%%%%%%%%%%%%
References~\cite{Campbell:2014kua} and~\cite{Jezo:2016ujg} introduce a generic method for interfacing \POWHEG{} processes that include radiation in decaying resonances with PS generators. 
According to this method, shower radiation from the resonance is left unrestricted, and a veto is applied \emph{a posteriori}: if any radiation in the decaying resonance shower is harder than the \POWHEG{} generated one, the event is discarded, and the same LH event is showered again. 
The practical implementation of the veto procedure depends on whether a dipole, as in \PythiaEightPtwo{}, or an angular-ordered shower, as in \HerwigSevenPone{}, is used.
If we are using a dipole ($\pt$-ordered) shower, it is sufficient to check the first shower-generated emission from the bottom quark and (if present at the LH level) from the gluon arising in top decay.
In the case of angular-ordered showers, as in \HerwigSevenPone{}, it is not enough to examine the first emission, because the hardest radiation may take place later.

Ref.~\cite{Ravasio:2018lzi} describes two on-the-fly veto procedures, one for \PythiaEightPtwo{} and one for \linebreak \HerwigSevenPone{}: each time an emission from a decayed top occurs, if its transverse momentum is larger than the one of the corresponding POWHEG emission, this emission is rejected and the shower evolution continues from the scale of the discarded emission.
In this work, we adapt and implement the same procedure in \PythiaSixPfour{} and \HerwigSixPfive{} in order to allow their matching to \bbfourl{} and \ttNLOdec{}.
%%%%%%%%%%%%%%%%%%%%%%%%%%%%%%%%%%%%%%%%%%%%%%%%%%%%%%%%%%%%%%%%%%%%%%%%%%%%%%%%%%%%%%%%%%%%%%%%%%%%%%%%%%%%%%%%%%%%%%%%%%%%%%%%%%%%%%%%%%%%%%%%%%%%

\section{Setup and methodology}
\label{sec:pheno}
%%%%%%%%%%%%%%%%%%%%%%%%%%%%%%%%%%%%%%%%%%%%%%%%%%%%%%%%%%%%%%%%%%%%%%%%%%%%%%%%%%%%%%%%%%%%%%%%%%%%%%%%%%%%%%%%%%%%%%%%%%%%%%%%%%%%%%%%%%%%%%%%%%%%
We simulate the process $p\,p \to b\, \bar{b}\,\fourl$, which is available in all three generators. 
It is dominated by top-pair production, with a smaller content of $Wt$ topologies. 
We perform our simulations for a center-of-mass energy of $\sqrt{s}=8$~TeV.  
We have used the {\tt MSTW2008nlo68cl} PDF set~\cite{Martin:2009iq} and, in the dominant resonance history, we have chosen as central renormalization and factorization scale the quantity $\mu$.
We define it as the geometric average of the transverse masses of the top and anti-top $\mu=\left[\left(E^2_t -p_{z,t}^2\right)\left(E^2_{\bar{t}}-p_{z,\bar{t}}^2\right)\right]^{1/4}$, where the top and anti-top energies $E_{t/\bar{t}}$ and longitudinal momenta $p_{z,t/\bar{t}}$ are evaluated at the underlying-Born level.
%%%%%%%%%%%%%%%%%%%%%%%%%%%%%%%%%%%%%%%%%%%%%%%%%%%%%%%%%%%%%%%%%%%%%%%%%%%%%%%%%%%%%%%%%%%%%%%%%%%%%%%%%%%%%%%%%%%%%%%%%%%%%%%%%%%%%%%%%%%%%%%%%%%%

%%%%%%%%%%%%%%%%%%%%%%%%%%%%%%%%%%%%%%%%%%%%%%%%%%%%%%%%%%%%%%%%%%%%%%%%%%%%%%%%%%%%%%%%%%%%%%%%%%%%%%%%%%%%%%%%%%%%%%%%%%%%%%%%%%%%%%%%%%%%%%%%%%%%
We make the $B$ hadrons stable, in order to simplify the definitions of $b$ jets.  
Jets are reconstructed using the {\tt Fastjet}~\cite{Cacciari:2011ma} implementation of the anti-$k_{\rm\sss T}$ algorithm~\cite{Cacciari:2008gp} with $R=0.5$.  
We denote as $B$~(${\bar B}$) the hardest (i.e.~largest \pT{}) $b$~($\bar{b}$) flavoured hadron. 
The $b$~($\bar{b}$) jet is the jet that contains the hardest $B$~($\bar{B}$).
It will be indicated as $j_B$~($j_{\bar{B}}$). 
We discard events where the $b$ jet and $\bar{b}$ jet coincide. 
The hardest $e^+$~($\mu^-$) and the hardest $\nu_e$~($\bar{\nu}_{\mu}$) are paired to reconstruct the $W^+$~($W^-$).  
The reconstructed top~(anti-top) quark is identified with the corresponding $W^+j_B$ ($W^-j_{\bar{B}}$) pair.  
In the following we will refer to the mass of this system as \mwbj{}.
We require the two $b$ jets to have $\pT>30\mbox{~GeV}$ and $|\eta|<2.5$.
These cuts suppress the single-top topologies. 
The hardest $e^+$ and the hardest $\mu^-$ must satisfy $\pT>20\mbox{~GeV}$ and $|\eta|<2.4$.
%%%%%%%%%%%%%%%%%%%%%%%%%%%%%%%%%%%%%%%%%%%%%%%%%%%%%%%%%%%%%%%%%%%%%%%%%%%%%%%%%%%%%%%%%%%%%%%%%%%%%%%%%%%%%%%%%%%%%%%%%%%%%%%%%%%%%%%%%%%%%%%%%%%%

%%%%%%%%%%%%%%%%%%%%%%%%%%%%%%%%%%%%%%%%%%%%%%%%%%%%%%%%%%%%%%%%%%%%%%%%%%%%%%%%%%%%%%%%%%%%%%%%%%%%%%%%%%%%%%%%%%%%%%%%%%%%%%%%%%%%%%%%%%%%%%%%%%%%
The reconstructed mass observable bears a nearly direct relation with the top mass. 
If two generators with the same $\mt$ input parameter yield a reconstructed mass peak position that differ by a certain amount, we can be certain that if they are used to extract the top mass they will yield results that differ by roughly the same amount in the opposite direction. 
%%%%%%%%%%%%%%%%%%%%%%%%%%%%%%%%%%%%%%%%%%%%%%%%%%%%%%%%%%%%%%%%%%%%%%%%%%%%%%%%%%%%%%%%%%%%%%%%%%%%%%%%%%%%%%%%%%%%%%%%%%%%%%%%%%%%%%%%%%%%%%%%%%%%

\section{Results}
\label{sec:mwbj}

%%%%%%%%%%%%%%%%%%%%%%%%%%%%%%%%%%%%%%%%%%%%%%%%%%%%%%%%%%%%%%%%%%%%%%%%%%%%%%%%%%%%%%%%%%%%%%%%%%%%%%%%%%%%%%%%%%%%%%%%%%%%%%%%%%%%%%%%%%%%%%%%%%%%
We begin by comparing the three generators, all interfaced to \PythiaEightPtwo{}, for our reference top-mass value of 172.5~GeV.
\begin{figure}[tb]
\centering
\includegraphics[width=0.35\textwidth]{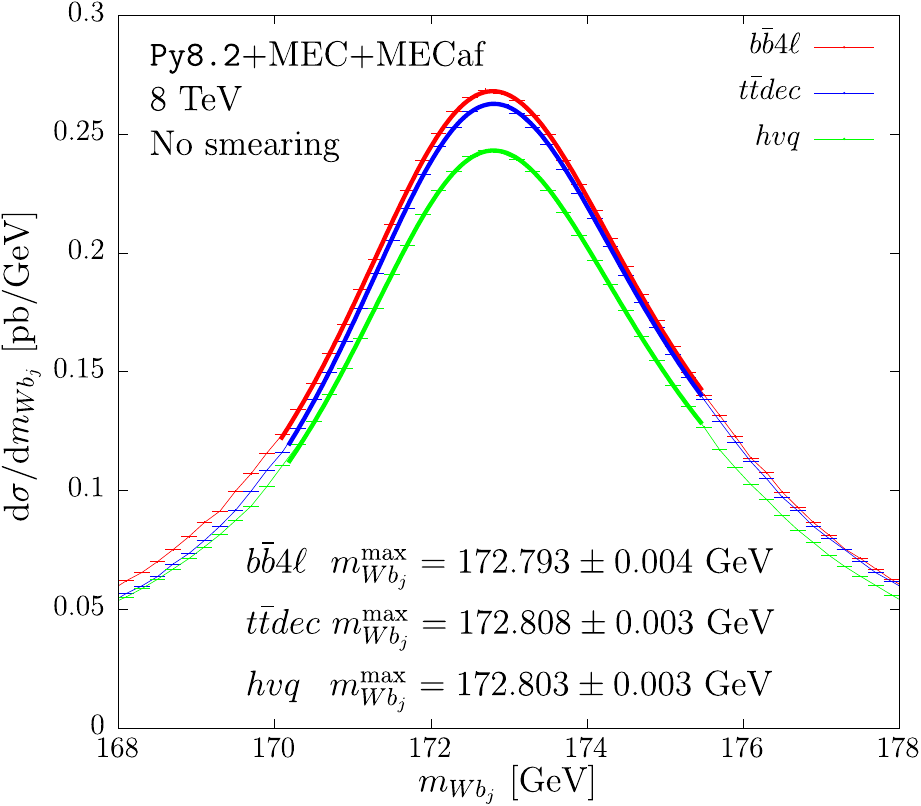}\hskip1.5cm
\includegraphics[width=0.35\textwidth]{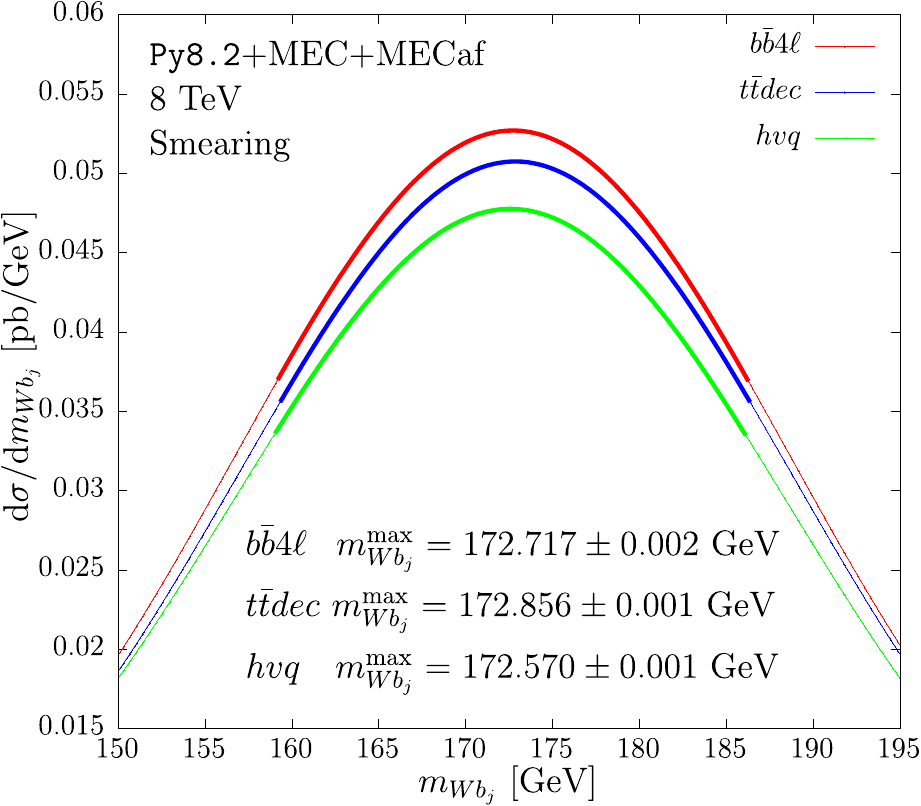}
\caption{${d\sigma}/{d \mwbj}$ distribution obtained with the \bbfourl{}, \ttNLOdec{} and \hvq{}
  generators interfaced with \PythiaEightPtwo{}, for $\mt=172.5$~GeV.}
\label{fig:MassPeaks-py8-bb4l-ttb-hvq-MEC}
\end{figure}
In the left panel of Fig.~\ref{fig:MassPeaks-py8-bb4l-ttb-hvq-MEC} we show the \mwbj{} distributions at the full hadron level in the case of an ideal detector.
In order to mimic experimental resolution effects, we smear our distribution with a Gaussian of width $\sigma=15$~GeV, \[f_{\rm smeared}(x) = \mathcal{N} \int dy f(y) \exp\left(-(y-x)^2/(2\sigma^2)\right)\,,\] where $\mathcal{N}$ is a normalization constant.
The smeared distributions are shown in the right panel of Fig.~\ref{fig:MassPeaks-py8-bb4l-ttb-hvq-MEC}.
Smearing effects are such that more importance is given to the region away from the peak, where there are larger differences between the generators.
%%%%%%%%%%%%%%%%%%%%%%%%%%%%%%%%%%%%%%%%%%%%%%%%%%%%%%%%%%%%%%%%%%%%%%%%%%%%%%%%%%%%%%%%%%%%%%%%%%%%%%%%%%%%%%%%%%%%%%%%%%%%%%%%%%%%%%%%%%%%%%%%%%%%

%%%%%%%%%%%%%%%%%%%%%%%%%%%%%%%%%%%%%%%%%%%%%%%%%%%%%%%%%%%%%%%%%%%%%%%%%%%%%%%%%%%%%%%%%%%%%%%%%%%%%%%%%%%%%%%%%%%%%%%%%%%%%%%%%%%%%%%%%%%%%%%%%%%%
We extract the position of the maximum by fitting the distribution with a skewed Lorentzian function of the form
$y(\mwbj)=b[1+d(\mwbj-a)]/\left[(\mwbj-a)^2+c^2\right]+e$.
The statistical error on the extracted $\mwbjmax$ is derived by propagating the errors of the parameters $a$, $b$, $c$, $d$ and $e$ in the expression of the peak.
%%%%%%%%%%%%%%%%%%%%%%%%%%%%%%%%%%%%%%%%%%%%%%%%%%%%%%%%%%%%%%%%%%%%%%%%%%%%%%%%%%%%%%%%%%%%%%%%%%%%%%%%%%%%%%%%%%%%%%%%%%%%%%%%%%%%%%%%%%%%%%%%%%%%

%%%%%%%%%%%%%%%%%%%%%%%%%%%%%%%%%%%%%%%%%%%%%%%%%%%%%%%%%%%%%%%%%%%%%%%%%%%%%%%%%%%%%%%%%%%%%%%%%%%%%%%%%%%%%%%%%%%%%%%%%%%%%%%%%%%%%%%%%%%%%%%%%%%%
Firstly we observe the \bbfourl{} and \ttNLOdec{} results are very close to each other, both in the shape and in the peak position.
We take this as an indication that interference effects in radiation and other off-shell effects, that are included in \bbfourl{} but not in \ttNLOdec{}, have a very minor impact on the peak position, at least if we consider a measurement with an ideal resolution.
We see a negligible difference in peak positions in the non-smeared case for all generators, while, in the smeared case, the \hvq{} generator differs from \bbfourl{} by \diffhvqbbfourl~MeV, similar in magnitude to the case of \ttbnlodec{}, but with the opposite sign.
%%%%%%%%%%%%%%%%%%%%%%%%%%%%%%%%%%%%%%%%%%%%%%%%%%%%%%%%%%%%%%%%%%%%%%%%%%%%%%%%%%%%%%%%%%%%%%%%%%%%%%%%%%%%%%%%%%%%%%%%%%%%%%%%%%%%%%%%%%%%%%%%%%%%

\begin{figure}[tb]
\centering
\includegraphics[width=0.35\textwidth]{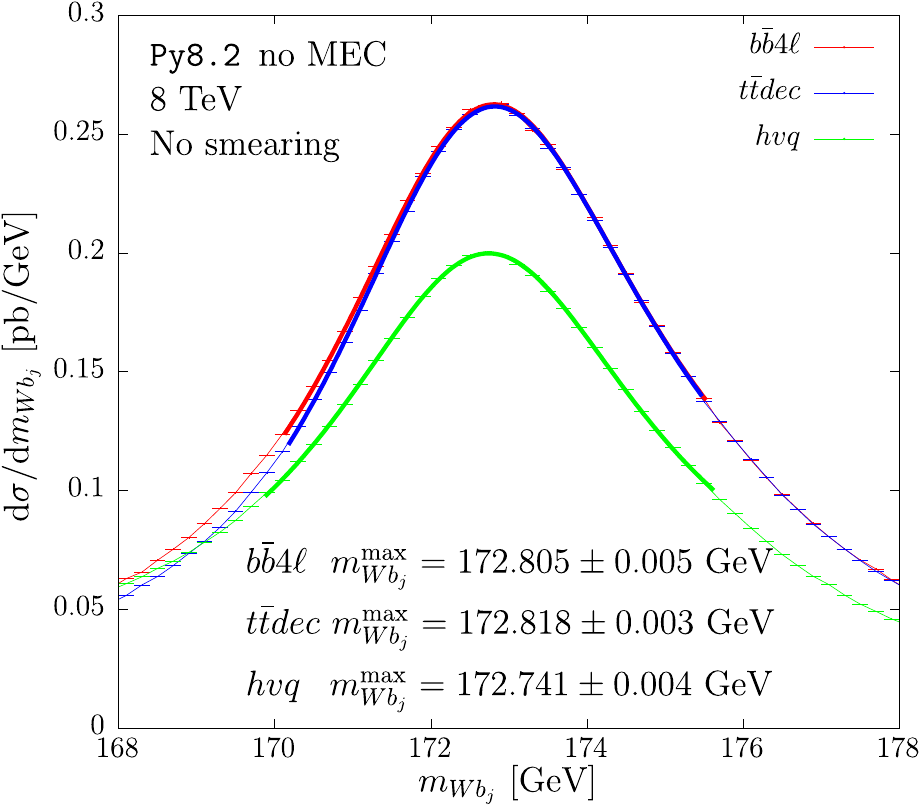}\hskip1.5cm
\includegraphics[width=0.35\textwidth]{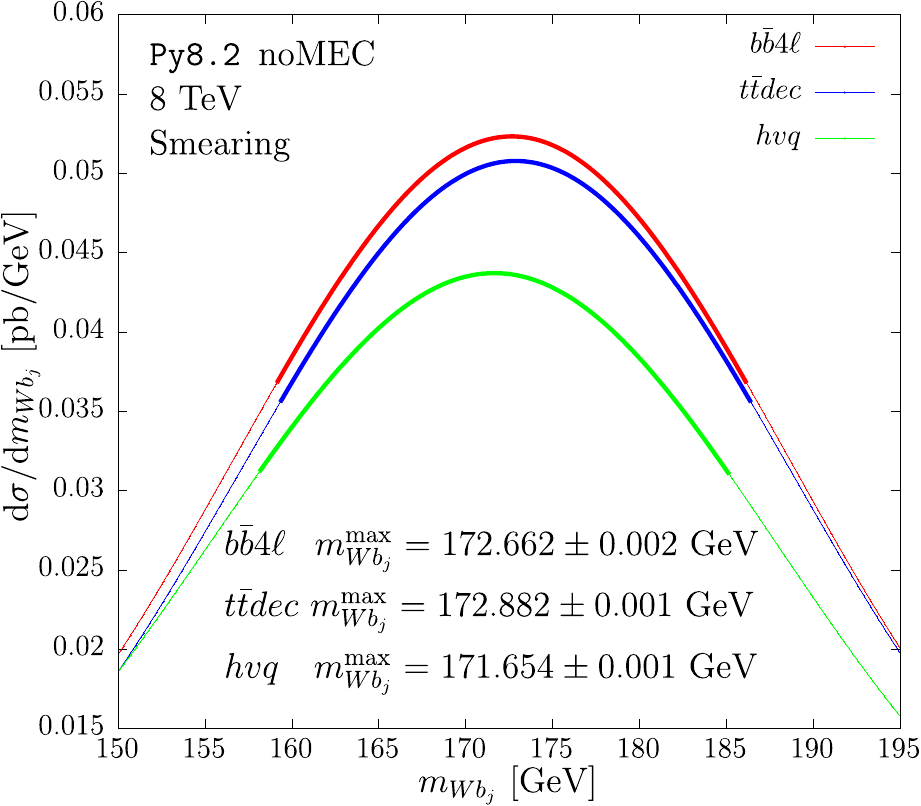}
\caption{Smeared ${d\sigma}/{d \mwbj}$ distribution obtained with the \bbfourl{}, \ttNLOdec{} and \hvq{} generators interfaced with \PythiaEightPtwo{}, for
  $\mt=172.5$~GeV.}
\label{fig:MassPeaks-py8-bb4l-ttb-hvq-NOMEC}
\end{figure}

%%%%%%%%%%%%%%%%%%%%%%%%%%%%%%%%%%%%%%%%%%%%%%%%%%%%%%%%%%%%%%%%%%%%%%%%%%%%%%%%%%%%%%%%%%%%%%%%%%%%%%%%%%%%%%%%%%%%%%%%%%%%%%%%%%%%%%%%%%%%%%%%%%%%
Moreover we notice that \hvq{}, in spite of the fact that it does not implement NLO corrections in top decay, yields distributions that are very close to those of the most accurate \bbfourl{} generator.
This is due to the fact that \PythiaEightPtwo{} includes MEC in top decays by default, which are equivalent, up to an irrelevant normalization factor, to NLO corrections.  
This observation is confirmed by examining, in Fig.~\ref{fig:MassPeaks-py8-bb4l-ttb-hvq-NOMEC}, 
the impact of the MEC setting on our predictions.
When MEC are switched off, we see a considerable shift, near 1~GeV, in the \hvq{} result for the peak position in the smeared distribution, and a very minor one in the \bbfourl{} and \ttbnlodec{} generators, that include the hardest emission off the $b$ quarks.
Thus we conclude that the MEC in \PythiaEightPtwo{} do a decent job in simulating top decay as far as the \mwbj{} distribution is concerned. 
The remaining uncertainty of roughly \diffaverage~MeV in the case of both \hvq{} and \ttbnlodec{} generators, pulling in opposite directions, is likely due to the approximate treatment of off-shell effects.
%%%%%%%%%%%%%%%%%%%%%%%%%%%%%%%%%%%%%%%%%%%%%%%%%%%%%%%%%%%%%%%%%%%%%%%%%%%%%%%%%%%%%%%%%%%%%%%%%%%%%%%%%%%%%%%%%%%%%%%%%%%%%%%%%%%%%%%%%%%%%%%%%%%%

%%%%%%%%%%%%%%%%%%%%%%%%%%%%%%%%%%%%%%%%%%%%%%%%%%%%%%%%%%%%%%%%%%%%%%%%%%%%%%%%%%%%%%%%%%%%%%%%%%%%%%%%%%%%%%%%%%%%%%%%%%%%%%%%%%%%%%%%%%%%%%%%%%%%
We continue by comparing the results obtained by interfacing \hvq{} and \bbfourl{} to two versions of the \Pythia{} and \Herwig{} SMCs.
In Ref.~\cite{Ravasio:2018lzi} we found that interfacing the \bbfourl{} generator to \HerwigSevenPone{} and \PythiaEightPtwo{} leads to large differences in the fitted peak position, at the full as well as at the shower only level (i.e.~with the hadronization and the MPI switched off).
For \hvq{} the differences are considerably smaller, although not quite negligible.
Furthermore, by varying the radius of jet reconstruction we find the two SMCs to behave differently, suggesting that at least one of the two may not describe the data fairly.
%%%%%%%%%%%%%%%%%%%%%%%%%%%%%%%%%%%%%%%%%%%%%%%%%%%%%%%%%%%%%%%%%%%%%%%%%%%%%%%%%%%%%%%%%%%%%%%%%%%%%%%%%%%%%%%%%%%%%%%%%%%%%%%%%%%%%%%%%%%%%%%%%%%%

\begin{figure*}[tb]
\centering
\includegraphics[width=0.35\textwidth]{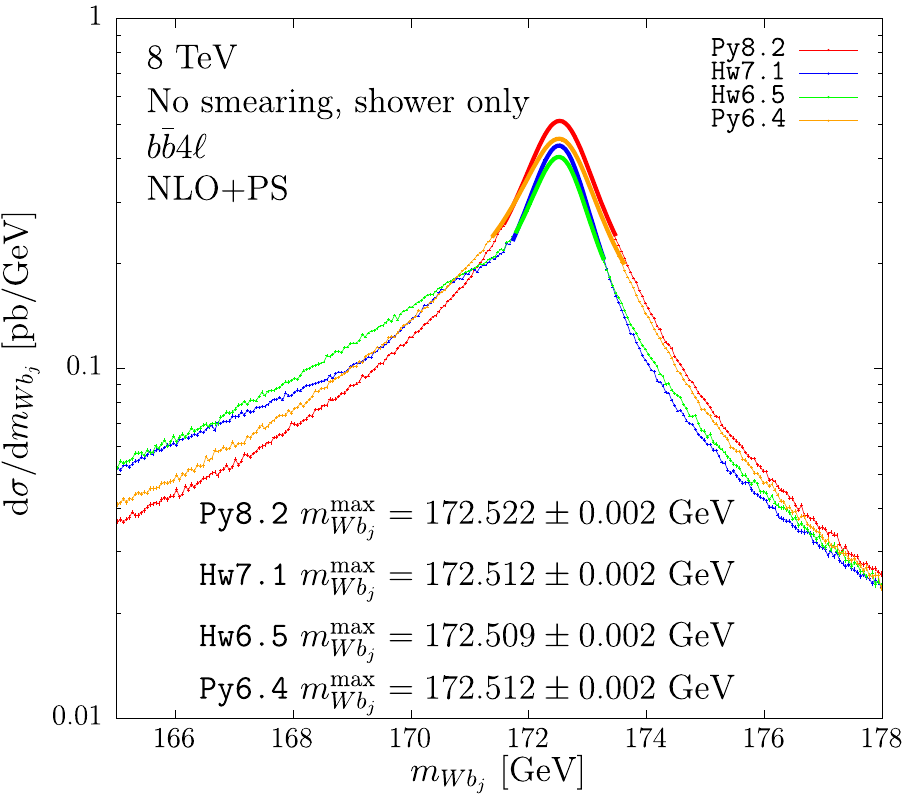}\hskip1.5cm
\includegraphics[width=0.35\textwidth]{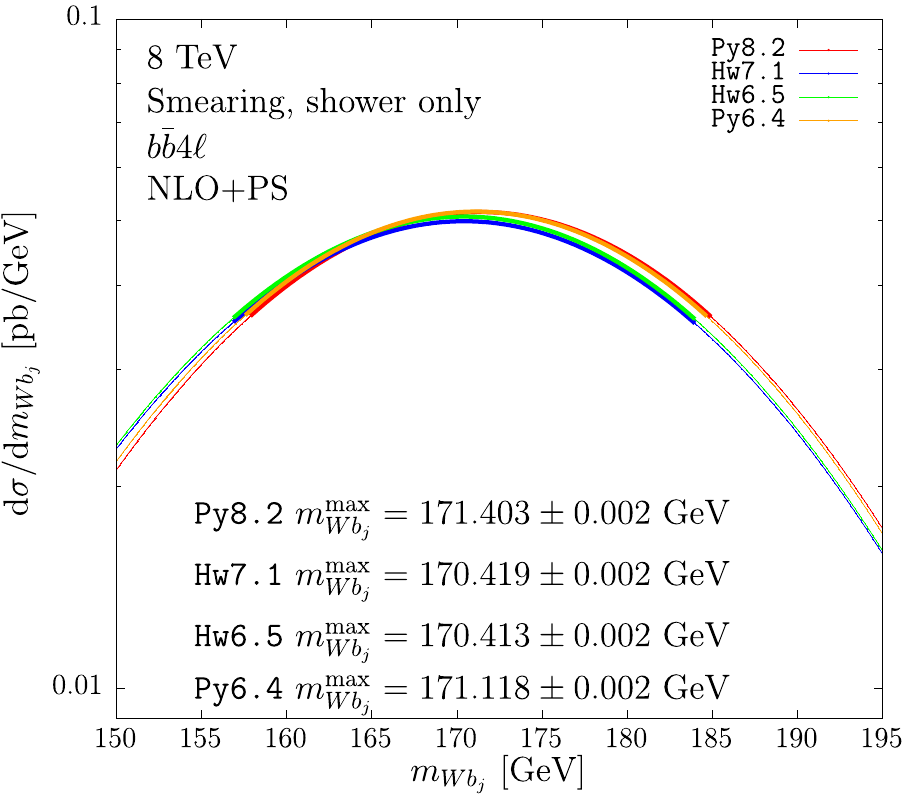}
  \caption{${d\sigma}/{d \mwbj}$ distribution obtained by showering the
    \bbfourl{} results with \PythiaEightPtwo{} and \HerwigSevenPone{},
    at parton-shower level~(left) and with hadronization and underlying events~(right).}
\label{fig:hwgvspythia}
\end{figure*}

%%%%%%%%%%%%%%%%%%%%%%%%%%%%%%%%%%%%%%%%%%%%%%%%%%%%%%%%%%%%%%%%%%%%%%%%%%%%%%%%%%%%%%%%%%%%%%%%%%%%%%%%%%%%%%%%%%%%%%%%%%%%%%%%%%%%%%%%%%%%%%%%%%%%
In Fig.~\ref{fig:hwgvspythia} we show the \mwbj{} distributions at the shower only level in the non-smeared and smeared cases, in the left and right panels respectively.
We find that the peak positions obtained using the two versions of \Herwig{} agree very well both in the ideal and the smeared case, while the peak positions of the two \Pythia{} versions differ by 285~MeV in the smeared case, as compared to $\sim$ 1~GeV difference when comparing results obtained with \Herwig{} and with \Pythia{}.
%%%%%%%%%%%%%%%%%%%%%%%%%%%%%%%%%%%%%%%%%%%%%%%%%%%%%%%%%%%%%%%%%%%%%%%%%%%%%%%%%%%%%%%%%%%%%%%%%%%%%%%%%%%%%%%%%%%%%%%%%%%%%%%%%%%%%%%%%%%%%%%%%%%%

\section{Summary}
%%%%%%%%%%%%%%%%%%%%%%%%%%%%%%%%%%%%%%%%%%%%%%%%%%%%%%%%%%%%%%%%%%%%%%%%%%%%%%%%%%%%%%%%%%%%%%%%%%%%%%%%%%%%%%%%%%%%%%%%%%%%%%%%%%%%%%%%%%%%%%%%%%%%
A thorough comparison of state of the art top-pair generators against those used in recent top-mass determinations reveals discrepancies.
Most of the observed differences in the peak position of the $Wb$-jet system's invariant mass are large enough to justify the use of the most accurate generators, but not to drastically overturn the conclusions of current measurements.
The differences due to the use of either \PythiaEightPtwo{} or \HerwigSevenPone{} to complete the shower evolution, however, extend beyond the current experimental errors on the top mass.
This pattern is confirmed by their previous versions, \PythiaSixPfour{} and \HerwigSixPfive{}.
If, as it seems, the differences found between \Pythia{} and \Herwig{} are due to the different shower models, the very minimal message that can be drawn from our work is that, in order to assess a meaningful theoretical error in top-mass measurements, the use of different shower models, associated with different NLO+PS generators, is mandatory.
Note, however, that our analysis is very crude and simple, and a realistic analysis, including an adequate tuning of the MC parameters and eventually also jet-energy calibration using hadronic $W$ decays in the same top events, performed by experimental collaborations could lead to an increased consistency between the \Pythia{} and \Herwig{} results.
%%%%%%%%%%%%%%%%%%%%%%%%%%%%%%%%%%%%%%%%%%%%%%%%%%%%%%%%%%%%%%%%%%%%%%%%%%%%%%%%%%%%%%%%%%%%%%%%%%%%%%%%%%%%%%%%%%%%%%%%%%%%%%%%%%%%%%%%%%%%%%%%%%%%

\section{Acknowledgments}
%%%%%%%%%%%%%%%%%%%%%%%%%%%%%%%%%%%%%%%%%%%%%%%%%%%%%%%%%%%%%%%%%%%%%%%%%%%%%%%%%%%%%%%%%%%%%%%%%%%%%%%%%%%%%%%%%%%%%%%%%%%%%%%%%%%%%%%%%%%%%%%%%%%%
We would like to thank B.R.~Webber for his kind help with a consistent implementation of the on-the-fly veto procedure in \HerwigSixPfive{}.
%%%%%%%%%%%%%%%%%%%%%%%%%%%%%%%%%%%%%%%%%%%%%%%%%%%%%%%%%%%%%%%%%%%%%%%%%%%%%%%%%%%%%%%%%%%%%%%%%%%%%%%%%%%%%%%%%%%%%%%%%%%%%%%%%%%%%%%%%%%%%%%%%%%%

\bibliography{paper}
\end{document}

%% file: texmacros.tex
\newcommand\POWHEGBOX{{\tt POWHEG\,BOX}}
\newcommand\POWHEG{{POWHEG}}
\newcommand\MCatNLO{{\tt MC@NLO}}
\newcommand\Pythia{{\tt Pythia}}
\newcommand\Herwig{{\tt Herwig}}
\newcommand\PythiaSixPfour{{\tt Pythia6.4}}

\newcommand\HerwigSixPfive{{\tt Herwig6.5}}

\newcommand\HerwigSevenPone{{\tt Herwig7.1}}

\newcommand\PythiaEightPtwo{{\tt Pythia8.2}}

\newcommand\RES{{\tt POWHEG\,BOX\,RES}}

\newcommand\sss{\mathchoice%
{\displaystyle}%
{\scriptstyle}%
{\scriptscriptstyle}%
{\scriptscriptstyle}%
}
\newcommand{\pt}{\ensuremath{p_{\sss T}}\xspace}

\def\beq{\begin{equation}}
\def\beqn{\begin{eqnarray}}
\def\eeq{\end{equation}}
\def\eeqn{\end{eqnarray}}

\def\({\left(} 
\def\){\right)} 
 
\newcommand     \MSB            {\ifmmode {\overline{\rm MS}} \else
                                 $\overline{\rm MS}$\fi}

\newcommand\mt{m_{t}}

\newcommand\fourl{e^+\nu_{\sss e}\, \mu^-\bar{\nu}_{\sss \mu}}

\newcommand\mwbj{\ensuremath{m_{Wb_j}}}
\newcommand\mwbjmax{\ensuremath{m_{Wb_j}^{\max}}}

\newcommand\pT{\ensuremath{p_{\sss\rm  T}}}

\newcount\minutes 
\newcount\scratch 
\def\timestamp{%
\scratch=\time 
\divide\scratch by 60 
\edef\hours{\the\scratch} 
\multiply\scratch by 60 
\minutes=\time 
\advance\minutes by -\scratch 
---$\,$\hours:\null 
\ifnum\minutes< 10 0\fi 
\the\minutes}

\definecolor{mygray}{gray}{0.5}

\newcommand\ttbnlodecPlot{$t\bar{t}dec$}
\newcommand\bbllllPlot{$b\bar{b}4\ell$}
\newcommand\hvqPlot{$hvq$}

\newcommand\hvq{\hvqPlot}
\newcommand\bbfourl{\bbllllPlot}
\newcommand\ttNLOdec{\ttbnlodecPlot}
\newcommand\ttbnlodec{\ttNLOdec}

%% file: global_definitions.tex
\newcommand\diffhvqbbfourl{$-147$}
\newcommand\diffaverage{$ 140$}